\documentclass[12pt]{article}
\usepackage{amsmath,amsfonts,feynmp}

\makeatletter\@addtoreset{equation}{section}\makeatother

\def\be{\begin{equation}}
\def\ee{\end{equation}}
\def\bea{\begin{eqnarray}}
\def\eea{\end{eqnarray}}

\makeatletter\@addtoreset{equation}{section}\makeatother

\renewcommand{\title}[1]{\vbox{\center\LARGE{#1}}\vspace{5mm}}
\renewcommand{\author}[1]{\vbox{\center#1}\vspace{5mm}}
\newcommand{\address}[1]{\vbox{\center\em#1}}
\newcommand{\email}[1]{\vbox{\center\tt#1}\vspace{5mm}}

\begin{document}

\unitlength = .8mm

\begin{titlepage}
\begin{center}
\hfill \\
\hfill \\
\vskip 1cm

\title{Spin Chains in ${\cal N}=6$ Superconformal Chern-Simons-Matter Theory}

\author{Davide Gaiotto$^{1,a}$,
Simone Giombi$^{2,b}$,
Xi Yin$^{2,c}$}

\address{$^1$Institute for Advanced Study, Princeton, NJ, USA\\
\medskip
$^2$Jefferson Physical Laboratory, Harvard University,\\
Cambridge, MA 02138 USA}

\email{$^a$dgaiotto@gmail.com,
$^b$giombi@physics.harvard.edu,
$^c$xiyin@fas.harvard.edu}

\end{center}

\abstract{
In this note we study spin chain operators in the ${\cal N}=6$ Chern-Simons-matter theory
recently proposed by Aharony, Bergman, Jafferis and Maldacena to be dual to type IIA string
theory in $AdS_4\times \mathbb{CP}^3$. We study the two-loop dilatation operator in the gauge theory,
and compare to the Penrose limit on the string theory side.
}

\vfill

\end{titlepage}

\eject \tableofcontents

\section{Introduction}

The long standing problem of finding an exact description of the CFT dual to
M-theory on $AdS_4\times S^7$ (and orbifolds thereof), or the low energy limit of the world volume theory
of $N$ coinciding M2-branes, was solved beautifully in a recent paper of
Aharony, Bergman, Jafferis and Maldacena \cite{Aharony:2008ug}.
The dual gauge theory is a special case of the ${\cal N}=3$ superconformal Chern-Simons-matter (CSM) theories
studied in \cite{Gaiotto:2007qi} (see \cite{Schwarz:2004yj,Kapustin:1994mt,Chen:1992ee,Avdeev:1992jt,Avdeev:1991za} 
for earlier works), which has quiver type matter content
and enhanced ${\cal N}=6$ supersymmetry.
In particular, the 't Hooft limit of the ${\cal N}=6$ CSM theory is argued to be dual to type IIA
string theory on $AdS_4\times \mathbb{CP}^3$. See also \cite{Benna:2008zy,
Minahan:2008hf,Imamura:2008nn,Honma:2008jd,Nishioka:2008gz,Bhattacharya:2008bj} for subsequent works on this
theory, and \cite{Bagger:2006sk,Gustavsson:2007vu,VanRaamsdonk:2008ft,
Distler:2008mk,Gomis:2008be,Ho:2008ei,Gustavsson:2008bf}
for recent works on M2-brane world volume theories.

In this paper we make a step toward understanding the details of the duality
between the ${\cal N}=6$ CSM theory and type IIA string on $AdS_4\times \mathbb{CP}^3$
in non(near)-BPS sectors, by exploring both spin chain operators in the superconformal gauge theory
(continuing on \cite{Gaiotto:2007qi}) and the Penrose limit of the string theory dual.
We study the two-loop dilatation operators in subsectors of the spin chain,
as well as the dispersion relation and scattering of impurities in an infinite chain that preserves
a centrally extended $SU(2|2)$ superconformal algebra.
The central charge of the $SU(2|2)$ algebra plays a key role in determining the exact dispersion
relations of the impurities. It is related to the momentum $P$ along the spin chain in the form
$$
Z = f(\lambda) (1-e^{2\pi i P})
$$
where $f(\lambda)$ is a nontrivial function of the 't Hooft coupling $\lambda=N/k$. We find that
$f(\lambda)$ scales differently with $\lambda$ at weak coupling (from perturbative gauge theory)
and at strong coupling (from the Penrose limit).
We discuss operator mixing and match multiplets in the weak coupling regime
with those in the pp-wave limit. We also present some preliminary discussions on the
giant magnons in $AdS_4\times\mathbb{CP}^3$.

Note added in proof: Upon completion of the bulk of this work, we received
\cite{Nishioka:2008gz} and \cite{Minahan:2008hf}, which contain results
that overlap with different parts of this paper.

\section{The ${\cal N}=6$ Chern-Simons-matter theory}

\subsection{Lagrangian}

It will be useful for us to formulate ${\cal N}=6$ Chern-Simons-matter theory in the ${\cal N}=2$ language.
The gauge group will be $U(N)\times U(N)$, with a pair of chiral fields $A_i$ ($i=1,2$) in the bifundamental
representation $({\bf N},{\bf \bar N})$, and $B_i$ in the conjugate representation
$({\bf \bar N},{\bf N})$. There is an ${\cal N}=2$ superpotential
\begin{equation}
W = {4\pi\over k} {\rm Tr} \left( A_1 B_1 A_2 B_2 - A_1 B_2 A_2 B_1 \right)
\end{equation}
This theory possesses ${\cal N}=6$ supersymmetry, and is exactly conformal, with superconformal group $OSp(6|4)$. The scalar
components of $(A_1, A_2, B_1^\dagger, B_2^\dagger)$ transform in the ${\bf 4}$ of $SU(4)_R$,
whereas $(B_1, B_2, A_1^\dagger, A_2^\dagger)$ transform in the ${\bf \bar4}$.

The scalar potential can be written as $V = V_D + V_F$, where $V_F=|\partial W/\partial A_i|^2
+|\partial W/\partial B_i|^2$, and $V_D$ comes the coupling of the scalar
fields to the auxiliary fields $\sigma$ and $\tilde\sigma$ (which lie in the ${\cal N}=2$ gauge multiplet
and take values in the adjoint of the two $U(N)$'s),
\begin{equation}
V_D = {\rm Tr} \left[(\sigma A_i-A_i\tilde\sigma) (A_i^\dagger\sigma -\tilde\sigma A_i^\dagger)\right]
+ {\rm Tr} \left[(\tilde\sigma B_i - B_i\sigma) (B_i^\dagger\tilde \sigma - \sigma B_i^\dagger)\right]
\end{equation}
where
\begin{equation}
\begin{aligned}
&\sigma = {2\pi\over k} (A_i A_i^\dagger - B_i^\dagger B_i),\\
&\tilde\sigma = -{2\pi\over k}(B_i B_i^\dagger - A_i^\dagger A_i).
\end{aligned}
\end{equation}
There are quartic boson-fermion coupling of the form
\begin{equation}
\begin{aligned}
{\cal L}_{F}&= {\cal L}_Y-{\rm Tr} \left(\psi_{A_i}^\dagger \sigma \psi_{A_i}- \psi_{A_i}^\dagger \psi_{A_i}\tilde\sigma\right)
-{\rm Tr}\left(\psi_{B_i}^\dagger \tilde\sigma \psi_{B_i}- \psi_{B_i}^\dagger \psi_{B_i}\sigma\right)
\\
&-{\rm Tr}\left(A_i^\dagger \chi^\dagger \psi_{A_i} - \tilde\chi^\dagger A_i^\dagger \psi_{A_i}+c.c.\right)
-{\rm Tr}\left(B_i^\dagger \tilde\chi^\dagger \psi_{B_i} - \chi^\dagger B_i^\dagger \psi_{B_i}+c.c.\right)
\end{aligned}
\end{equation}
where $\chi$ and $\tilde\chi$ are fermionic auxiliary fields in the ${\cal N}=2$ gauge multiplet,
\begin{equation}
\begin{aligned}
&\chi = {2\pi\over k} (\psi_{A_i} A_i^\dagger - B_i^\dagger \psi_{B_i}),\\
&\tilde\chi = -{2\pi\over k}(\psi_{B_i} B_i^\dagger - A_i^\dagger \psi_{A_i}).
\end{aligned}
\end{equation}
and ${\cal L}_Y$ is the Yukawa coupling,
\begin{equation}
\begin{aligned}
{\cal L}_Y &= {\partial^2 W\over \partial \phi_i \partial \phi_j } \psi_i \psi_j + c.c.\\
&= {4\pi\over k} {\rm Tr} (A_1 B_1 \psi_{A_2}\psi_{B_2} + \cdots)
\end{aligned}
\end{equation}

\subsection{Supersymmetry transformations}
\label{susy}

In manifestly ${\cal N}=6$ supersymmetric notation, we can write the supercharges as $Q_{IJ}=(Q^{IJ})^\dagger={1\over 2}\epsilon_{IJKL}
\bar Q^{KL}$, where $I,J,K,L=1,\cdots,4$. The scalars and fermions are denoted by $\phi_I$, $\bar\phi^I$,
$(\psi_I)_\alpha$ and $\bar\psi^I_\alpha$. One can explicitly identify them with the components fields
of ${\cal N}=2$ chiral multiplets as
\begin{equation}
\begin{aligned}
&\phi_1 = A_1,~~~\phi_2 = A_2,~~~\phi_3=B_1^\dagger, ~~~\phi_4=B_2^\dagger,\\
&\psi_1 = -\psi_{A_2}^\dagger,~~~\psi_2 = \psi_{A_1}^\dagger, ~~~
\psi_3 = -\psi_{B_2},~~~\psi_4 = \psi_{B_1}.
\end{aligned}
\end{equation}
The action of the supercharges on the fields is as follows
\begin{equation}\label{susygen}
\begin{aligned}
& Q_{IJ} \phi_K = \epsilon_{IJKL} \bar\psi^L,\\
& Q_{IJ} \bar\phi^K = \delta_I^K\psi_J - \delta_J^K \psi_I,\\
& (Q_{IJ})_\alpha (\psi_K)_\beta = \epsilon_{IJKL} i\sigma^\mu_{\alpha\beta} D_\mu \bar\phi^L
+{2\pi i\over k}\epsilon_{\alpha\beta} \epsilon_{IJKL}( \bar\phi^L \phi_M \bar\phi^M-\bar\phi^M \phi_M \bar\phi^L)
+{4\pi i\over k} \epsilon_{\alpha\beta}\epsilon_{IJLM} \bar\phi^L \phi_K \bar\phi^M,\\
& (Q_{IJ})_\alpha (\bar\psi^K)_\beta = \delta_I^K \left[i\sigma^\mu_{\alpha\beta} D_\mu \phi_J
-{2\pi i\over k}\epsilon_{\alpha\beta} ( \phi_J \bar\phi^M \phi_M-\phi_M \bar\phi^M \phi_J)\right]\\
& ~~~~~~ - \delta_J^K \left[i\sigma^\mu_{\alpha\beta} D_\mu \phi_I
-{2\pi i\over k}\epsilon_{\alpha\beta} ( \phi_I \bar\phi^M \phi_M-\phi_M \bar\phi^M \phi_I)\right]
-{4\pi i\over k} \epsilon_{\alpha\beta}(\phi_I \bar\phi^K \phi_J-\phi_J \bar\phi^K \phi_I),\\
& Q_{IJ} A_\mu = i\sigma_\mu \chi_{IJ}= {2\pi i\over k}\sigma_\mu(\phi_{[I} \psi_{J]} + {1\over 2} \epsilon_{IJKL} \bar\psi^{K}\bar\phi^{L}),\\
& Q_{IJ} \tilde A_\mu = i\sigma_\mu \tilde\chi_{IJ}= {2\pi i\over k}\sigma_\mu(\psi_{[J}\phi_{I]} + {1\over 2} \epsilon_{IJKL} \bar\phi^{L}\bar\psi^{K}).
\end{aligned}
\end{equation}

\section{Spin chains in ${\cal N}=6$ CS}

\subsection{$SU(2)_A\times SU(2)_B$ sector}

Let us focus on the $SU(2)_A\times SU(2)_B\times U(1)$ subgroup of $SU(4)_R$, where $A_i$ transform in the representation $({\bf 2},{\bf 1},+1)$, and $B_i$ in the representation $({\bf 1},{\bf 2},-1)$. Consider spin chains
of the form
\begin{equation}
{\rm Tr} (A_{i_1} B_1 A_{i_2} B_1 A_{i_3} B_1\cdots)
\end{equation}
These are chiral operators, but in general not primaries due to the superpotential. At two-loop,
the sextic scalar potential coming from the superpotential contributes to the anomalous dimension of
the above operator. The relevant potential term is
\begin{equation}
{16\pi^2\over k^2}{\rm Tr} \left[(A_1 B_1 A_2-A_2 B_1 A_1) (A_1 B_1 A_2-A_2 B_1 A_1)^\dagger\right]
\end{equation}
The potential terms in $V_D$ does not contribute at two-loop.
Similarly, the terms coupling the scalars to fermions in ${\cal L}_F$ do not have the right structure to
contribute to the two-loop anomalous dimension of the chiral operator either (other than an overall
shift which is fixed by the BPS bound for the chiral primaries, i.e. the operators with all the $A_i$'s
symmetrized).

\bigskip
\bigskip

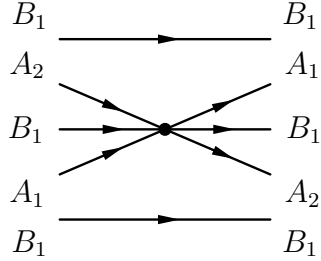
\centerline{\begin{fmffile}{diag1}
        \begin{tabular}{c}
            \begin{fmfgraph*}(35,30)
                \fmfstraight
                \fmfset{arrow_len}{.3cm}\fmfset{arrow_ang}{12}
                \fmfleft{i1,i2,i3,i4,i5}
                \fmflabel{$B_1$}{i1}
                \fmflabel{$A_1$}{i2}
                \fmflabel{$B_1$}{i3}
                \fmflabel{$A_2$}{i4}
                \fmflabel{$B_1$}{i5}
                \fmflabel{$B_1$}{o1}
                \fmflabel{$A_2$}{o2}
                \fmflabel{$B_1$}{o3}
                \fmflabel{$A_1$}{o4}
                \fmflabel{$B_1$}{o5}
                \fmfright{o1,o2,o3,o4,o5}
                \fmf{fermion}{i1,o1}
                \fmf{fermion}{i2,v1}
                \fmf{fermion}{v1,o2}
                \fmf{fermion}{i3,v1}
                \fmf{fermion}{v1,o3}
                \fmf{fermion}{i4,v1}
                \fmf{fermion}{v1,o4}
                \fmf{fermion}{i5,o5}
                \fmfdot{v1}
            \end{fmfgraph*}
        \end{tabular}
        \end{fmffile}
}
\bigskip
\noindent
The two-loop integral in the above diagram is
\begin{equation}
\int {d^3y\over (4\pi)^6} {1\over |y|^3 |x-y|^3} \sim {1\over 8\pi^2}{\ln\Lambda\over (4\pi|x|)^3}
\end{equation}
where $1/(4\pi |x|)$ is the scalar propagator in position space. There is also a factor of $16\pi^2\lambda^2$
from the vertices and contraction of color indices, and a factor of $1/2$ since we were calculating the
two point function of the spin chain operator as opposed to the anomalous dimension. Putting these together,
we then find the two-loop
spin chain Hamiltonian
\begin{equation}\label{XXXH}
H = -\lambda^2 \sum_i(P_{i,i+1}-1)
\end{equation}
This is the Hamiltonian of the Heisenberg XXX spin-$1/2$ chain. The dispersion relation of an impurity
in this $SU(2)$ sector moving with momentum $p$ is
\begin{equation}\label{disptwo}
E=4\lambda^2 \sin^2(\pi p) + {\cal O}(\lambda^3)
\end{equation}
There may be a regularization scheme dependent order $\lambda^3$ term, but its structure is the same as the
$\lambda^2$, since the corresponding three loop diagrams are obtained by attaching gauge propagators to
the two-loop diagrams.

Now let us allow the $B_1$'s to change into $B_2$ as well, so that the spin chain takes the form
\begin{equation}
{\rm Tr} (A_{i_1} B_{j_1} A_{i_2} B_{j_2} A_{i_3} B_{j_3}\cdots)
\end{equation}
Once again, at two-loop the only potential term that contributes to the anomalous dimension
are the (${\cal N}=2$) F-terms. Furthermore, the exchanges of $A_1$ and $A_2$ across $B_1$ or $B_2$ have
the same amplitude, and similarly for the exchange of $B_1$ and $B_2$ across $A_i$. Therefore, we find that
at two-loop the $SU(2)_A\times SU(2)_B$ spin chain is two decoupled XXX spin-$1/2$ chains (of $A$'s and $B$'s respectively).

\subsection{The $SU(2|2)$ infinite chain}

To gain further insight we shall consider the infinite chain (the ``vacuum")
\begin{equation}
{\rm Tr} (A_1 B_1 A_1 B_1 A_1 B_1\cdots)
\label{vacuum}
\end{equation}
It preserves an $SU(2|2)$ subgroup of $OSp(6|4)$. The bosonic part of $SU(2|2)$ is $SU(2)_G\times
SU(2)_r\times U(1)_D$, where $SU(2)_G$ rotates $A_2, B_2^\dagger$ as a doublet, $SU(2)_r$ is
the rotation group in spacetime, and $U(1)_D$ is generated by $D$, defined to be
the anomalous dimension. More precisely, $D=\Delta -J$, where $\Delta$ is the conformal dimension and $J$ is the eigenvalue of the Cartan generator of $SU(2)_{G'}$, which is the group rotating $A_1, B_1^\dagger$ (and similarly $\psi_1,\psi_3$) as a doublet 
\footnote{Choosing the ``vacuum'' (\ref{vacuum}), one considers the breaking 
$SU(4)_R \rightarrow SU(2)_{G'} \times SU(2)_{G} \times U(1)$, and the vacuum preserves
$SU(2)_G\times U(1)$.
The extra $U(1)$, which assigns charge $+1$ to $A_1, B_1^\dagger$ and charge $-1$ to
$A_2, B_2^\dagger$, commutes with the generators of $SU(2|2)$.}. 
Therefore one has $J(A_1)=J(B_1)=\frac{1}{2}$,$\, J(A_2)=J(B_2)=0$, 
and similarly for the fermions. The odd generators of $SU(2|2)$ are denoted by $Q_{A\alpha}, \bar S_{A\alpha}$,
where $A$ is an $SU(2)_G$ doublet index, and $\alpha$ is the spacetime spinor index. The
superalgebra is
\begin{equation}\label{sutt}
\begin{aligned}
&\{Q_{A\alpha}, Q_{B\beta}\}= \epsilon_{AB} \epsilon_{\alpha\beta}
Z,~~~~\{\bar S_{A\alpha}, \bar S_{B\beta}\}= \epsilon_{AB} \epsilon_{\alpha\beta} \bar Z,\\
&\{Q_{A\alpha}, \bar S_{B\beta}\} = \epsilon_{AB} \epsilon_{\alpha\beta} D
+\epsilon_{AB} J_{\alpha\beta} + \epsilon_{\alpha\beta} T_{AB}.
\end{aligned}
\end{equation}
where $Z$ is a central charge, related to the momentum of the impurities in the infinite chain,
to be determined later.

Comparing with the supersymmetry transformations (\ref{susygen}), the pair of supercharges that preserve
the vacuum spin chain is
$(Q_{12},-Q_{14})\sim Q_A$. In particular,
$J(Q_A)=\frac{1}{2}$, and $D=\Delta-J$ commutes with the supercharges as required by the $SU(2|2)$ algebra.

The basic impurities are $A_2, B_2^\dagger, (\psi_{B_2}^\dagger)_\alpha$ in place of $A_1$, and similarly
$A_2^\dagger, B_2, (\psi_{A_2}^\dagger)_\alpha$ in place of $B_1$. At zero momentum they
transform in the minimal short representation of $SU(2|2)$. We will write $\phi_A = (A_2, B_2^\dagger)
=(\phi_2,\phi_4)$,
and $\chi_\alpha = (\psi_{B_2}^\dagger)_\alpha$. From (\ref{susygen}) we have the supersymmetry transformations
on $(\phi_A,\chi_\alpha)$
\begin{equation}\label{susya}
\begin{aligned}
&Q_{A\alpha} \phi_B \sim \epsilon_{AB} \chi_\alpha,\\
&Q_{A\alpha} \chi_\beta \sim \epsilon_{\alpha\beta} {2\pi i\over k}(\phi_A B_1 A_1 - A_1 B_1\phi_A).
\end{aligned}
\end{equation}
In terms of impurities with momentum $p$,
we have
\begin{equation}
\begin{aligned}
&Q_{A\alpha} |\phi_B(p)\rangle \sim \epsilon_{AB} |\chi_\alpha(p)\rangle,\\
&Q_{A\alpha} |\chi_\beta(p)\rangle \sim \epsilon_{\alpha\beta}{2\pi i\over k}(1-e^{2\pi ip}) |\phi_A(p)\rangle.
\end{aligned}
\end{equation}
In (\ref{sutt}) we have normalized $Q_A$ and $\bar S_A$ to be complex conjugates of one another in radial
quantization. In general they are related to the supercharges in (\ref{susygen})
by a rescaling, which a priori may depend on the coupling $\lambda$ due to quantum corrections
to $\bar S_A$. The central charge of the $SU(2|2)$ algebra takes the form $Z=f(\lambda)(1-e^{2\pi ip})$,
where $f(\lambda)$ is an undetermined function of $\lambda$.
The basic impurities (4 bosonic and 4 fermionic) fall into two short representations:
\begin{equation}
\begin{aligned}
\{[1,0]|[0,1]\} \oplus \{[1,0]|[0,1]\}
\end{aligned}
\end{equation}
We will call them $(2|2)_A$ and $(2|2)_B$ impurities for short.
The short multiplet saturates
the BPS bound \cite{Beisert:2005tm,Beisert:2006qh},
\begin{equation}\label{bound}
\Delta-J = D = \sqrt{{1\over 4}+4f(\lambda)^2 \sin^2(\pi p)}
\end{equation}
By comparison with the two-loop spin chain Hamiltonian (\ref{XXXH}), we determine that $f(\lambda)\simeq \lambda$
in the weak 't Hooft coupling limit.
Let us check this relation for the fermionic impurity $\psi_{B_2}^\dagger$. There is in fact only one
diagram allowed by the index structure that contributes to the exchange of $A_1$ with $\psi_{B_2}^\dagger$
across a $B_1$ along the chain, as follows.

\bigskip
\bigskip

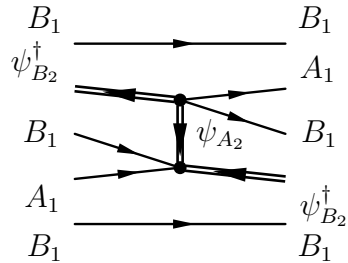
\centerline{       \begin{fmffile}{diag2}
        \begin{tabular}{c}
            \begin{fmfgraph*}(35,30)
                \fmfstraight
                \fmfset{arrow_len}{.3cm}\fmfset{arrow_ang}{12}
                \fmfleft{i1,i2,i3,i4,i5}
                \fmflabel{$B_1$}{i1}
                \fmflabel{$A_1$}{i2}
                \fmflabel{$B_1$}{i3}
                \fmflabel{$\psi_{B_2}^\dagger$}{i4}
                \fmflabel{$B_1$}{i5}
                \fmflabel{$B_1$}{o1}
                \fmflabel{$\psi_{B_2}^\dagger$}{o2}
                \fmflabel{$B_1$}{o3}
                \fmflabel{$A_1$}{o4}
                \fmflabel{$B_1$}{o5}
                \fmfright{o1,o2,o3,o4,o5}
                \fmf{fermion}{i1,o1}
                \fmf{fermion}{i2,v1}
                \fmf{fermion}{i3,v1}
                \fmf{fermion}{v2,o3}
                \fmf{fermion}{v2,o4}
                \fmf{fermion}{i5,o5}
                \fmf{dbl_plain_arrow,tension=2}{v2,i4}
                \fmf{dbl_plain_arrow,tension=0,lab.side=left,lab=$\psi_{A_2}$}{v2,v1}
                \fmf{dbl_plain_arrow,tension=2}{o2,v1}
                \fmfdot{v1,v2}
            \end{fmfgraph*}
        \end{tabular}
        \end{fmffile} }
\bigskip
There is a factor of $16\pi^2\lambda^2$ coming from the F-term vertices,
and a factor $1/2$ to convert to the anomalous dimension. The fermion propagator in position space is
$i{\slash \!\!\!x}/(4\pi |x|^3)$.
The loop integral involved is
\begin{equation}
-i\int {d^3y d^3z\over (4\pi)^7} {(\slash \!\!\!x-\slash \!\!\!z)
(\slash \!\!\!z-\slash \!\!\!y)\slash \!\!\!y\over z^2 (x-y)^2 |y|^3 |x-z|^3 |y-z|^3}
\end{equation}
whose logarithmically divergent part is
\begin{equation}
\begin{aligned}
&-2i{\slash \!\!\!x\over |x|^5} \int {d^3y d^3z \over (4\pi)^7} {(z-y)\cdot y\over z^2 |y|^3 |y-z|^3}
= -2i{\slash \!\!\!x\over |x|^5} \int {d^3y \over (4\pi)^7|y|^3} y^\mu {\partial\over \partial y^\mu}\int d^3z  {1\over z^2 |y-z|}
\\
&={1\over 8\pi^2} {i\slash \!\!\!x \ln\Lambda\over (4\pi)^3|x|^5}
\end{aligned}
\end{equation}
The resulting anomalous dimension is identical to that of the $A_2$ (or $B_2^\dagger$) impurity,
which is expected since they are in the same short multiplet.

\subsection{Scattering and bound states}

\subsubsection{$(2|2)_A\otimes (2|2)_A$ sector}

Now let us consider the scattering of a pair of basic $(2|2)$ impurities, working
perturbatively at two-loop. First consider a pair of impurities both in the $(2|2)_A$ multiplet
(or similarly, both in the $(2|2)_B$ multiplet), consisting of the fields
$(A_2, B_2^\dagger; \psi_{B_2}^\dagger)$. In particular, two $A_2$ impurities
with momenta $p_1$ and $p_2$ scatter according to the Hamiltonian (\ref{XXXH}),
and can form a bound state with dispersion relation \cite{Faddeev:1996iy}
\begin{equation}
\Delta-J -1= 2\lambda^2 \sin^2(\pi p).
\end{equation}
This saturates the BPS bound for the $(4|4)$ short multiplet of spin content
$\{[2,0],[0,0]|[1,1]\}$ under $SU(2)_G\times SU(2)_r\subset SU(2|2)$. The bosonic part of this short multiplet consists
of the bound states of the pairs
\begin{equation}
A_2 A_2, ~B_2^\dagger B_2^\dagger,~A_2 B_2^\dagger,~\epsilon^{\alpha\beta}
(\psi_{B_2}^\dagger)_\alpha(\psi_{B_2}^\dagger)_\beta
\end{equation}
moving with momentum $p$. The wave function decays exponentially as the pair is separated along the chain.
There is another $(4|4)$ ``multiplet" of asymptotic scattering states of
two $(2|2)_A$ multiplets, of spin content $\{[0,2],[0,0]|[1,1]\}$, whose bosonic part consists of
\begin{equation}\label{scs}
|A_2(p_1)B_2(p_2)^\dagger-B_2(p_1)^\dagger A_2(p_2)^\dagger\rangle,~~~
\sigma_\mu^{\alpha\beta}
|(\psi_{B_2}^\dagger)_\alpha(p_1)(\psi_{B_2}^\dagger)_\beta(p_2)\rangle
\end{equation}
However, they cannot form bound states at two loop. This is easiest to see from the scattering
of a bosonic $(2|2)_A$ impurity, say $A_2$, with a fermionic $(2|2)_A$ impurity $\psi_{B_2}^\dagger$.
There is an exchange amplitude between $A_2$ and $\psi_{B_2}^\dagger$ (or $B_2$ and $\psi_{B_2}^\dagger$), as in the diagrams below,
which allows only one bound state between them of given total momentum $p$. This bound state is
already included in the $\{[2,0],[0,0]|[1,1]\}$ multiplet, and hence there are no fermionic bound states
to pair up with potential bound states coming from (\ref{scs}).

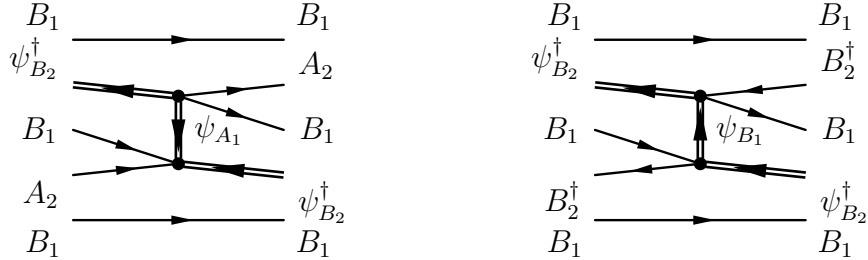
\bigskip
\bigskip

\centerline{        \begin{fmffile}{diag3}
        \begin{tabular}{c}
            \begin{fmfgraph*}(35,30)
                \fmfstraight
                \fmfset{arrow_len}{.3cm}\fmfset{arrow_ang}{12}
                \fmfleft{i1,i2,i3,i4,i5}
                \fmflabel{$B_1$}{i1}
                \fmflabel{$A_2$}{i2}
                \fmflabel{$B_1$}{i3}
                \fmflabel{$\psi_{B_2}^\dagger$}{i4}
                \fmflabel{$B_1$}{i5}
                \fmflabel{$B_1$}{o1}
                \fmflabel{$\psi_{B_2}^\dagger$}{o2}
                \fmflabel{$B_1$}{o3}
                \fmflabel{$A_2$}{o4}
                \fmflabel{$B_1$}{o5}
                \fmfright{o1,o2,o3,o4,o5}
                \fmf{fermion}{i1,o1}
                \fmf{fermion}{i2,v1}
                \fmf{fermion}{i3,v1}
                \fmf{fermion}{v2,o3}
                \fmf{fermion}{v2,o4}
                \fmf{fermion}{i5,o5}
                \fmf{dbl_plain_arrow,tension=2}{v2,i4}
                \fmf{dbl_plain_arrow,tension=0,lab.side=left,lab=$\psi_{A_1}$}{v2,v1}
                \fmf{dbl_plain_arrow,tension=2}{o2,v1}
                \fmfdot{v1,v2}
            \end{fmfgraph*}
        \end{tabular}
        \end{fmffile}
        ~~~~~~~~~~~~~~~~~~~~~~~\begin{fmffile}{diag4}
        \begin{tabular}{c}
            \begin{fmfgraph*}(35,30)
                \fmfstraight
                \fmfset{arrow_len}{.3cm}\fmfset{arrow_ang}{12}
                \fmfleft{i1,i2,i3,i4,i5}
                \fmflabel{$B_1$}{i1}
                \fmflabel{$B_2^\dagger$}{i2}
                \fmflabel{$B_1$}{i3}
                \fmflabel{$\psi_{B_2}^\dagger$}{i4}
                \fmflabel{$B_1$}{i5}
                \fmflabel{$B_1$}{o1}
                \fmflabel{$\psi_{B_2}^\dagger$}{o2}
                \fmflabel{$B_1$}{o3}
                \fmflabel{$B_2^\dagger$}{o4}
                \fmflabel{$B_1$}{o5}
                \fmfright{o1,o2,o3,o4,o5}
                \fmf{fermion}{i1,o1}
                \fmf{fermion}{v1,i2}
                \fmf{fermion}{i3,v1}
                \fmf{fermion}{v2,o3}
                \fmf{fermion}{o4,v2}
                \fmf{fermion}{i5,o5}
                \fmf{dbl_plain_arrow,tension=2}{v2,i4}
                \fmf{dbl_plain_arrow,tension=0,lab.side=right,lab=$\psi_{B_1}$}{v1,v2}
                \fmf{dbl_plain_arrow,tension=2}{o2,v1}
                \fmfdot{v1,v2}
            \end{fmfgraph*}
        \end{tabular}
        \end{fmffile}}

\bigskip

\noindent
It is plausible that the $(4|4)$ bound state of a pair of $(2|2)_A$ impurities, which we denote by
$(4|4)_A$, remains a short multiplet at strong coupling.

\subsubsection{$(2|2)_A\otimes (2|2)_B$ sector}

From the earlier discussion on $SU(2)_A\times SU(2)_B$ sector of the spin chain we know that
the $A_2$ and $B_2$ impurities from $(2|2)_A$ and $(2|2)_B$ do not interact at two-loop (but
this is not necessarily the case for other pairs of impurities in $(2|2)_A\otimes (2|2)_B$).
In particular we have two-impurity states, with $A_2$ of momentum $p_1$ and $B_2$ of momentum $p_2$,
denoted by $|A_2(p_1)B_2(p_2)\rangle$, which are eigenstates of the two-loop dilatation operator.
By $SU(2|2)$ symmetry, there must be a 16-dimensional long multiplet of threshold (non-)scattering
states, of spin content $\{[2,0],[0,2],[0,0],[0,0]|[1,1],[1,1]\}$. The $[2,0]$ part consists of
the scalar triplet
\begin{equation}\label{trip}
|A_2(p_1)B_2(p_2)\rangle,~~|B_2^\dagger(p_1) A_2^\dagger(p_2)\rangle,~~
|A_2(p_1) A_2^\dagger(p_2) - B_2^\dagger(p_1) B_2(p_2)\rangle.
\end{equation}
The product representation
$(2|2)_A\otimes (2|2)_B$ also consists of the $SU(2)_r$ triplet of fermion bilinear
$$
\sigma_\mu^{(\alpha\beta)}|(\psi_{B_2})_\alpha^\dagger(p_1) (\psi_{A_2})_\beta^\dagger(p_2)\rangle
$$
Naively one may expect this to be the $[0,2]$ part of the long multiplet.
However, these pairs of basic impurities are interacting at two-loop, and the corresponding
states are not eigenstates of the Hamiltonian. In particular, the exchange amplitude
$$\sigma_\mu^{(\alpha\beta)}|\cdots(\psi_{B_2})_\alpha^\dagger (\psi_{A_2})_\beta^\dagger A_1\cdots\rangle
\to \sigma_\mu^{(\alpha\beta)}|\cdots A_1(\psi_{A_2})_\beta^\dagger(\psi_{B_2})_\alpha^\dagger\cdots\rangle$$
vanishes at two-loop, as the above diagram vanishes when the spinor indices
$\alpha,\beta$ are symmetrized. This effect
leads to a repulsive contact (i.e. nearest neighbor) interaction between the impurities
$(\psi_{B_2})_\alpha^\dagger(p_1)$ and $(\psi_{A_2})_\beta^\dagger(p_2)$ in the $SU(2)_r$ triplet
sector.

The resolution to this seeming puzzle is due to operator mixing, between say $\sigma_\mu^{(\alpha\beta)}|(\psi_{B_2})_\alpha^\dagger(p_1) (\psi_{A_2})_\beta^\dagger(p_2)\rangle$ and $|D_\mu(p)\rangle$, the state of an impurity $D_\mu A_1$
or $D_\mu B_1$ moving at momentum $p=p_1+p_2$. At two-loop this can be computed from the
amplitude
$$\sigma_\mu^{(\alpha\beta)}|\cdots(\psi_{B_2})_\alpha^\dagger (\psi_{A_2})_\beta^\dagger \cdots\rangle
\to |\cdots A_1 D_\mu B_1\cdots\rangle$$
via the following Feynman diagrams

\bigskip
\bigskip

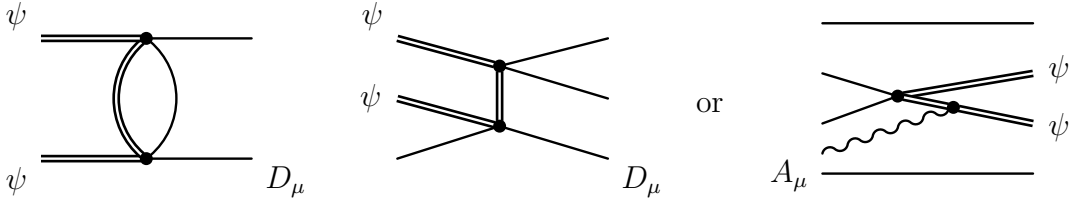
\centerline{\begin{fmffile}{diag6}
        \begin{tabular}{c}
            \begin{fmfgraph*}(35,20)
                \fmfstraight
                \fmfset{arrow_len}{.3cm}\fmfset{arrow_ang}{12}
                \fmfleft{i1,i2}
                \fmfright{o1,o2}
                \fmflabel{$\psi$}{i1}
                \fmflabel{$\psi$}{i2}
                \fmflabel{$D_\mu$}{o1}
                \fmf{plain}{v1,o1}
                \fmf{plain}{v2,o2}
                \fmf{dbl_plain}{v1,i1}
                \fmf{dbl_plain}{v2,i2}
                \fmf{plain,right=.5,tension=0}{v1,v2}
                \fmf{dbl_plain,left=.5,tension=0}{v1,v2}
                \fmfdot{v1,v2}
            \end{fmfgraph*}
        \end{tabular}
        \end{fmffile}~~~~~~~
        \begin{fmffile}{diag7}
        \begin{tabular}{c}
            \begin{fmfgraph*}(35,20)
                \fmfstraight
                \fmfset{arrow_len}{.3cm}\fmfset{arrow_ang}{12}
                \fmfleft{i1,i2,i3}
                \fmfright{o1,o2,o3}
                \fmffixed{(0.85h,-.23h)}{i2,v1}
                \fmffixed{(0.85h,-.23h)}{i3,v2}
                \fmflabel{$\psi$}{i2}
                \fmflabel{$\psi$}{i3}
                \fmflabel{$D_\mu$}{o1}
                \fmf{plain,tension=1}{v1,o1}
                \fmf{plain,tension=0}{v2,o2}
                \fmf{plain}{v2,o3}
                \fmf{plain,tension=0.3}{v1,i1}
                \fmf{dbl_plain}{v1,i2}
                \fmf{dbl_plain}{v2,i3}
                \fmf{dbl_plain,tension=.6}{v1,v2}
                \fmfdot{v1,v2}
            \end{fmfgraph*}
        \end{tabular}
        \end{fmffile}~~~~~~or~~~~~
        \begin{fmffile}{diag8}
        \begin{tabular}{c}
            \begin{fmfgraph*}(35,25)
                \fmfstraight
                \fmfset{arrow_len}{.3cm}\fmfset{arrow_ang}{12}
                \fmfright{i1,i3,i4,u1}
                \fmfleft{o1,o3,o4,w1}
                \fmffixed{(0,0.2h)}{i2,i3}
                \fmffixed{(0,0.2h)}{o2,o3}
                \fmflabel{$\psi$}{i3}
                \fmflabel{$\psi$}{i4}
                \fmflabel{$A_\mu$}{o2}
                \fmf{plain,tension=1}{v1,o3}
                \fmf{plain,tension=1}{v1,o4}
                \fmf{dbl_plain,tension=.7}{v2,i3}
                \fmf{dbl_plain,tension=.7}{v1,i4}
                \fmf{dbl_plain,tension=1}{v1,v2}
                \fmf{wiggly,tension=0}{o2,v2}
                \fmf{plain}{o1,i1}
                \fmf{plain}{u1,w1}
                \fmfdot{v1,v2}
            \end{fmfgraph*}
        \end{tabular}
        \end{fmffile}
            }

\bigskip

\noindent It will turn out that we can determine the coefficients of these amplitudes simply based on the
consistency requirement that there are threshold non-scattering states of such mixed operators.

A simple example of such mixing at zero momentum (more precisely, at momentum $p=1$) is the following protected
operator obtained by acting on the vacuum chain with supercharges,
\begin{equation}\label{prot}
\begin{aligned}
&(Q_{13})_\alpha(Q_{24})_\beta |A_1B_1A_1B_1\cdots\rangle = -\sum_{n=even,\;m=odd} |
(\psi_{B_2})_\beta^\dagger(n) (\psi_{A_2})_\alpha^\dagger(m)\rangle\\
&~~~~~+i\sigma^\mu_{\alpha\beta}\sum_{n~even} |D_\mu A_1(n)\rangle
-i\epsilon_{\alpha\beta} \left(\sum_{n~odd}|\sigma_{24}(n)\rangle
-\sum_{n~even}|\tilde\sigma_{24}(n)\rangle\right) \\
&(Q_{24})_\beta(Q_{13})_\alpha |A_1B_1A_1B_1\cdots\rangle = \sum_{n=even,\;m=odd} |
(\psi_{B_2})_\beta^\dagger(n) (\psi_{A_2})_\alpha^\dagger(m)\rangle\\
&~~~~~+i\sigma^\mu_{\alpha\beta}\sum_{n~odd} |D_\mu B_1(n)\rangle
+i\epsilon_{\alpha\beta} \left(\sum_{n~odd}|\sigma_{24}(n)\rangle-\sum_{n~even}|\tilde\sigma_{24}(n)\rangle
\right) \\
\end{aligned}
\end{equation}
where $\sigma_{24}$ and $\tilde\sigma_{24}$ are defined as
\begin{equation}
\begin{aligned}
& \sigma_{24} = {2\pi \over k}(-\bar\phi^1\phi_1+\bar\phi^2\phi_2-\bar\phi^3\phi_3+\bar\phi^4\phi_4),\\
& \tilde\sigma_{24} = {2\pi \over k}(-\phi_1\bar\phi^1+\phi_2\bar\phi^2-\phi_3\bar\phi^3+\phi_4\bar\phi^4).
\end{aligned}
\end{equation}
The sum of the two lines in (\ref{prot}) gives the total derivative of the vacuum chain, whereas the difference
gives another protected operator (in both the $SU(2)_r$ triplet and singlet sector). A special case is when the length of the chain is 2,
and we obtain a component of the $SU(4)_R$ current
\begin{equation}
{(J_\mu)_1}^3={\rm Tr}\left[A_1 D_\mu B_1-(D_\mu A_1) B_1 -i (\psi_{B_2})_\alpha^\dagger(p_1)
\sigma_\mu^{\alpha\beta}(\psi_{A_2})_\beta^\dagger(p_2)\right]
\end{equation}

Let us now compute the operator mixing in the sector of a $(\psi_{B_2})_\alpha^\dagger$ and a
$(\psi_{A_2})_\beta^\dagger$ impurity in the triplet of $SU(2)_r$, or a single impurity $D_\mu$.
Denote by $|n,m\rangle$ the state with $(\psi_{B_2})_\alpha^\dagger$ at position $2n$
and $(\psi_{A_2})_\beta^\dagger$ at position $2m+1$, with the spinor indices contracted
by $\sigma_\mu^{\alpha\beta}$. Denote by $|D(n)\rangle$ the state of a $D_\mu$ acting on the
site $2n$, and by $|D'(n)\rangle$ the state of $D_\mu$ acting on the site $2n+1$. The two-loop dilatation operator then acts on these states as
\begin{equation}
\begin{aligned}
&H |n,m\rangle = -\lambda^2 (|n-1,m\rangle + |n+1,m\rangle + |n,m-1\rangle+|n,m+1\rangle -
4|n,m\rangle),\\
&~~~~~ ~~~~~~~~~~~~~~~~~~~~~~~~~~~~~~~~~~~~~~~~~~~~~~~~~~~~~~~~~~~~~~~~~~~~~~
(|n-m-{1\over 2}|>{1\over 2}) \\
&H |n,n\rangle = -\lambda^2 (|n-1,n\rangle + |n,n+1\rangle -
(4-\delta)|n,n\rangle)\\
&~~~~~~~~~~~~~~~ - c_1 \lambda^2 (|D(n)\rangle - |D'(n)\rangle )
-c_2\lambda^2 (|D'(n-1)\rangle - |D(n+1)\rangle ), \\
&H |n,n-1\rangle = -\lambda^2 (|n+1,n-1\rangle + |n,n-2\rangle -
(4-\delta)|n,n-1\rangle)\\
&~~~~~~~~~~~~~~~- c_1 \lambda^2 (|D'(n-1)\rangle - |D(n)\rangle )
-c_2\lambda^2 (|D(n-1)\rangle - |D'(n)\rangle ), \\
& H|D(n)\rangle = -c_1\lambda^2 ( |n,n\rangle-|n,n-1\rangle)
-c_2\lambda^2 (|n+1,n\rangle-|n-1,n-1\rangle)
\\
&~~~~~~~~~~~~~~~ - c_3\lambda^2 (|D'(n-1)\rangle+|D'(n)\rangle-2|D(n)\rangle),\\
&~~~~~~~~~~~~~~~ - c_4\lambda^2 (|D(n-1)\rangle+|D(n+1)\rangle-2|D(n)\rangle),\\
& H|D'(n)\rangle = -c_1\lambda^2 ( |n+1,n\rangle-|n,n\rangle)
-c_2\lambda^2 (|n+1,n+1\rangle-|n,n-1\rangle)\\
&~~~~~~~~~~~~~~~- c_3\lambda^2 (|D(n)\rangle+|D(n+1)\rangle-2|D'(n)\rangle)\\
&~~~~~~~~~~~~~~~ - c_4\lambda^2 (|D'(n-1)\rangle+|D'(n+1)\rangle-2|D'(n)\rangle).
\end{aligned}
\end{equation}
where $\delta, c_1, c_2, c_3, c_4$ are constants that can be computed from the two-loop diagrams.
In particular, $\delta$ is a potentially nonzero correction to the anomalous dimension when
$(\psi_{B_2})_\alpha^\dagger$ and $(\psi_{A_2})_\beta^\dagger$ are next to each other (although
it will turn out to be zero in this case). The
coefficients $c_1$ and $c_2$ are due to the mixing between adjacent fermion pair $(\psi_{B_2})_\alpha^\dagger(\psi_{A_2})_\beta^\dagger$ and a $D_\mu$ on the nearest and
next-to-nearest neigboring sites, respectively. $c_3$ is due to the mixing of a $D_\mu$ impurity
with another $D_\mu$ on nearest
neighboring sites, according to the diagrams

\bigskip
\bigskip

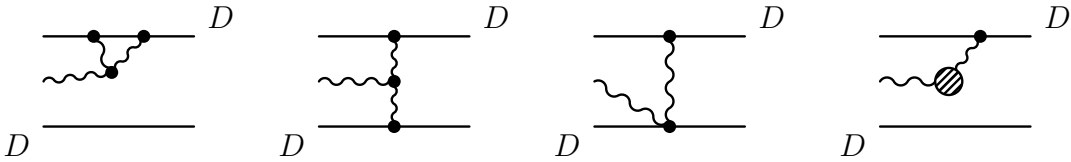
\centerline{\begin{fmffile}{diag9}
        \begin{tabular}{c}
            \begin{fmfgraph*}(25,15)
                \fmfstraight
                \fmfset{arrow_len}{.3cm}\fmfset{arrow_ang}{12}
                \fmffixed{(0.2h,-0.4h)}{v2,v1}
                \fmfleft{i3,i2,i1}
                \fmfright{o3,o2,o1}
                \fmflabel{$D$}{i3}
                \fmflabel{$D$}{o1}
                \fmf{plain}{i1,v2}
                \fmf{plain}{v2,v3}
                \fmf{plain}{v3,o1}
                \fmf{plain}{i3,o3}
                \fmf{photon,tension=0}{i2,v1}
                \fmf{photon,tension=0}{v2,v1}
                \fmf{photon,tension=0}{v3,v1}
                \fmfdot{v1,v2,v3}
            \end{fmfgraph*}
        \end{tabular}
        \end{fmffile}~~~~~
        \begin{fmffile}{diag10}
        \begin{tabular}{c}
            \begin{fmfgraph*}(25,15)
                \fmfstraight
                \fmfset{arrow_len}{.3cm}\fmfset{arrow_ang}{12}
                \fmffixed{(0,0.5h)}{v2,v1}
                \fmfleft{i1,i2,i3}
                \fmfright{o1,o2,o3}
                \fmflabel{$D$}{i1}
                \fmflabel{$D$}{o3}
                \fmf{plain}{i1,v2}
                \fmf{plain}{v2,o1}
                \fmf{plain}{i3,v3}
                \fmf{plain}{v3,o3}
                \fmf{photon,tension=0}{i2,v1}
                \fmf{photon,tension=0}{v2,v1}
                \fmf{photon,tension=0}{v3,v1}
                \fmfdot{v1,v2,v3}
            \end{fmfgraph*}
        \end{tabular}
        \end{fmffile}~~~~~
        \begin{fmffile}{diag11}
        \begin{tabular}{c}
            \begin{fmfgraph*}(25,15)
                \fmfstraight
                \fmfset{arrow_len}{.3cm}\fmfset{arrow_ang}{12}
                %\fmffixed{(0,0.5h)}{v2,v1}
                \fmfleft{i1,i2,i3}
                \fmfright{o1,o2,o3}
                \fmflabel{$D$}{i1}
                \fmflabel{$D$}{o3}
                \fmf{plain}{i1,v1}
                \fmf{plain}{v1,o1}
                \fmf{plain}{i3,v2}
                \fmf{plain}{v2,o3}
                \fmf{photon,tension=0}{i2,v1}
                \fmf{photon,tension=0}{v2,v1}
                \fmfdot{v1,v2}
            \end{fmfgraph*}
        \end{tabular}
        \end{fmffile}
        ~~~~~
        \begin{fmffile}{diag12}
        \begin{tabular}{c}
            \begin{fmfgraph*}(25,15)
                \fmfstraight
                \fmfset{arrow_len}{.3cm}\fmfset{arrow_ang}{12}
                \fmffixed{(0.35h,0.5h)}{v2,v1}
                \fmfleft{i1,i2,i3}
                \fmfright{o1,o2,o3}
                \fmflabel{$D$}{i1}
                \fmflabel{$D$}{o3}
                \fmf{plain}{i3,v1}
                \fmf{plain,tension=2}{v1,o3}
                \fmf{plain}{i1,o1}
                \fmfv{d.sh=circle,d.fi=shaded,d.si=.3h}{v2}
                \fmf{photon,tension=0}{i2,v2}
                \fmf{photon,tension=0}{v2,v1}
                \fmfdot{v1}
            \end{fmfgraph*}
        \end{tabular}
        \end{fmffile}
            }

\bigskip

\noindent whereas $c_4$ is due to mixing of $D_\mu$'s on next-to-nearest neighboring sites, from the following diagrams
\bigskip
\bigskip

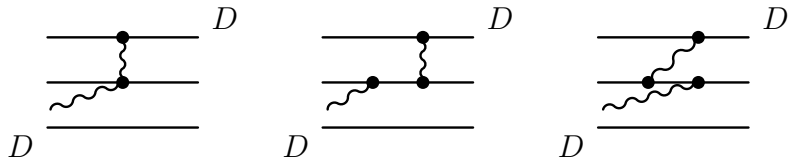
\centerline{\begin{fmffile}{diag13}
        \begin{tabular}{c}
            \begin{fmfgraph*}(25,15)
                \fmfstraight
                \fmfset{arrow_len}{.3cm}\fmfset{arrow_ang}{12}
                \fmffixed{(.8h,.3h)}{v3,v1}
                \fmfleft{i1,i2,i3}
                \fmfright{o1,o2,o3}
                \fmflabel{$D$}{i1}
                \fmflabel{$D$}{o3}
                \fmf{plain}{i1,o1}
                \fmf{plain}{i2,v1,o2}
                \fmf{plain}{i3,v2,o3}
                \fmf{photon,tension=0}{v3,v1,v2}
                \fmfdot{v1,v2}
            \end{fmfgraph*}
        \end{tabular}
        \end{fmffile}~~~~~
        \begin{fmffile}{diag14}
        \begin{tabular}{c}
            \begin{fmfgraph*}(25,15)
                \fmfstraight
                \fmfset{arrow_len}{.3cm}\fmfset{arrow_ang}{12}
                \fmffixed{(.5h,.3h)}{v3,v1}
                \fmfleft{i1,i2,i3}
                \fmfright{o1,o2,o3}
                \fmflabel{$D$}{i1}
                \fmflabel{$D$}{o3}
                \fmf{plain}{i1,o1}
                \fmf{plain}{i2,v1,v1a,o2}
                \fmf{plain}{i3,v2}
                \fmf{plain,tension=2}{v2,o3}
                \fmf{photon,tension=0}{v3,v1}
                \fmf{photon,tension=0}{v1a,v2}
                \fmfdot{v1,v2,v1a}
            \end{fmfgraph*}
        \end{tabular}
        \end{fmffile}~~~~~
        \begin{fmffile}{diag15}
        \begin{tabular}{c}
            \begin{fmfgraph*}(25,15)
                \fmfstraight
                \fmfset{arrow_len}{.3cm}\fmfset{arrow_ang}{12}
                \fmffixed{(.5h,.3h)}{v3,v1}
                \fmfleft{i1,i2,i3}
                \fmfright{o1,o2,o3}
                \fmflabel{$D$}{i1}
                \fmflabel{$D$}{o3}
                \fmf{plain}{i1,o1}
                \fmf{plain}{i2,v1,v1a,o2}
                \fmf{plain}{i3,v2}
                \fmf{plain,tension=2}{v2,o3}
                \fmf{photon,tension=0}{v3,v1a}
                \fmf{photon,tension=0}{v1,v2}
                \fmfdot{v1,v2,v1a}
            \end{fmfgraph*}
        \end{tabular}
        \end{fmffile}
            }

\bigskip

\noindent We will not compute these diagrams directly, but simply determine them from the existence of the
threshold states at general momenta below. The result is
\begin{equation}
\delta=0,~~~ c_1= {1\over \sqrt{2}},~~~c_2= -{1\over \sqrt{2}},~~~c_3=1,~~~c_4={1\over 2}.
\end{equation}
A general state of total momentum $p$ takes the form
\begin{equation}
|\Psi\rangle=\sum_{n,m} e^{\pi ip(n+m+{1\over 2})} f(n-m) |n,m\rangle
+ g\sum_n e^{2\pi ipn} |D(n)\rangle + g'\sum_n e^{2\pi ip(n+{1\over 2})} |D'(n)\rangle
\end{equation}
Suppose $|\Psi\rangle$ is an energy eigenstate $H|\Psi\rangle = \lambda^2 E|\Psi\rangle$.
This is equivalent to the equations
\begin{equation}\label{threeq}
\begin{aligned}
&2 \cos(\pi p) (f(n-1)+f(n+1))-4 f(n) = -Ef(n),~~~~~n\geq 2{~\rm or ~}n\leq -1 \\
&2\cos(\pi p) f(2)-4f(1)+\sqrt{2}\cos(\pi p) (-e^{-\pi i p/2}g+e^{\pi i p/2} g') = -E f(1),\\
&2\cos(\pi p) f(-1)-4f(0)+\sqrt{2}\cos(\pi p) (-e^{-\pi i p/2}g'+e^{\pi i p/2} g) = -E f(0),\\
& (\cos(2\pi p)-1)g+2(\cos(\pi p)g'-g)+\sqrt{2}\cos(\pi p) (-e^{\pi i p/2} f(1)+e^{-\pi i p/2} f(0))=-Eg,\\
& (\cos(2\pi p)-1)g'+2(\cos(\pi p)g'-g)+\sqrt{2}\cos(\pi p) (-e^{\pi i p/2} f(0)+e^{-\pi i p/2} f(1))=-Eg',\\
\end{aligned}
\end{equation}
The threshold state is given by
\begin{equation}
\begin{aligned}
&f(n) = -{i\over \sqrt{2}}e^{-i(n-1/2) \alpha} (\cos(\pi p)-e^{-i\alpha}),~~~~(n\geq 1)\\
&f(n) = -{i\over \sqrt{2}}e^{-i(n-1/2) \alpha} (\cos(\pi p)-e^{i\alpha}),~~~~(n\leq 0)\\
&g=\sin({\pi p-\alpha\over 2}),~~~~g'=\sin({\pi p+\alpha\over 2}),\\
&E = 4\sin^2({\pi p-\alpha\over 2})+4\sin^2({\pi p+\alpha\over 2}).
\end{aligned}
\end{equation}
where $\alpha=\pi(p_1-p_2)$ is the difference between the momenta of the two $\psi$ impurities.
In particular the protected operators in the triplet sector of (\ref{prot}) are given by the special case
$p=1$, $\alpha=0$. One can also check that there are no bound states at two loop.\footnote{An attempt
to find such bound states is to set say $e^{i\alpha}=\cos(\pi p)$ with purely imaginary $\alpha$
in (\ref{threeq}), but this state is growing as opposed to decaying, exponentially, in
the separation between $\psi_{B_2}^\dagger$ and $\psi_{A_2}^\dagger$.
} A priori these threshold states may not survive at higher loops, but they may survive in
the pp-wave limit as unbound $(2|2)_A$ and $(2|2)_B$ impurities.

There is also operator mixing in the $[0,0]$ part of the long multiplet. For instance, the fermion bilinear singlet
\begin{equation}\label{sing}
|\psi_{B_2}^\dagger(p_1) \psi_{A_2}^\dagger(p_2)\rangle=\epsilon^{\alpha\beta}|(\psi_{B_2})_\alpha^\dagger(p_1) (\psi_{A_2})_\beta^\dagger(p_2)\rangle
\end{equation}
can mix with four bosons, via diagrams such as the following

\bigskip
\bigskip
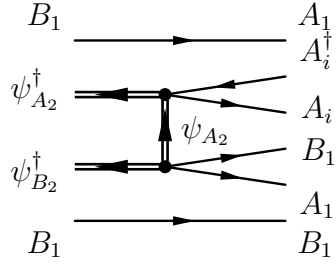
\centerline{\begin{fmffile}{diag5}
        \begin{tabular}{c}
            \begin{fmfgraph*}(35,30)
                \fmfstraight
                \fmfset{arrow_len}{.3cm}\fmfset{arrow_ang}{12}
                \fmfleft{i1,i2,i3,i4,i5,i6}
                \fmflabel{$B_1$}{i1}
                \fmffixed{(0h,0.4h)}{v1,v2}
                \fmffixed{(0h,-0.1h)}{s2,i2}
                \fmffixed{(0h,0.1h)}{s5,i5}
                \fmffixed{(0.5h,0h)}{s2,v1}
                \fmffixed{(0.5h,0h)}{s5,v2}
                \fmflabel{$\psi_{B_2}^\dagger$}{s2}
                \fmflabel{$\psi_{A_2}^\dagger$}{s5}
                \fmflabel{$B_1$}{i6}
                \fmflabel{$B_1$}{o1}
                \fmflabel{$A_1$}{o2}
                \fmflabel{$B_1$}{o3}
                \fmflabel{$A_i$}{o4}
                \fmflabel{$A_i^\dagger$}{o5}
                \fmflabel{$A_1$}{o6}
                \fmfright{o1,o2,o3,o4,o5,o6}
                \fmf{fermion}{i1,o1}
                \fmf{fermion}{i6,o6}
                \fmf{dbl_plain_arrow}{v1,s2}
                \fmf{dbl_plain_arrow,lab=$\psi_{A_2}$}{v1,v2}
                \fmf{dbl_plain_arrow}{v2,s5}
                \fmf{fermion,tesion=.2}{v1,o2}
                \fmf{fermion,tesion=.2}{v1,o3}
                \fmf{fermion,tesion=.2}{o5,v2,o4}
                \fmfdot{v1,v2}
            \end{fmfgraph*}
        \end{tabular}
        \end{fmffile}}
\bigskip
\noindent We have seen this mixing at zero momentum already in (\ref{prot}).

\section{Penrose limit of type IIA string theory on $AdS_4\times {\mathbb{CP}}^3$}

The 't Hooft limit of the ${\cal N}=6$ superconformal Chern-Simons-matter theory is dual
to type IIA string theory on $AdS_4\times {\mathbb{CP}}^3$ \cite{Aharony:2008ug}.
The metric on $AdS_4\times {\mathbb{CP}}^3$ can be written as \cite{Pope:1984bd}
\begin{equation}\label{cpmetric}
\begin{aligned}
ds^2 &= R^2\left\{-\cosh^2\rho dt^2 + d\rho^2 + \sinh^2\rho d\Omega_2^2 \right. \\ &\left.+4 d\mu^2+ 4\sin^2\mu \left[d\alpha^2 +
{1\over 4}\sin^2\alpha(\sigma_1^2+\sigma_2^2+\cos^2\alpha\sigma_3^2)+{1\over 4}\cos^2\mu (d\chi+\sin^2\alpha \sigma_3)^2 \right]\right\}
\end{aligned}
\end{equation}
Here $R$ is the radius of the $AdS_4$, and $\sigma_{1,2,3}$ are left-invariant 1-forms on an $S^3$, parameterized by $(\theta,\phi,\psi)$,
\begin{equation}
\begin{aligned}
& \sigma_1 = \cos\psi d\theta + \sin \psi \sin\theta d\phi,\\
& \sigma_2 = \sin\psi d\theta - \cos\psi \sin\theta d\phi,\\
& \sigma_3 = d\psi+\cos\theta d\phi.
\end{aligned}
\end{equation}
The range of the coordinates is $0\le \mu,\alpha \le \pi/2\,,0\le \theta \le \pi\,,0\le\phi\le 2\pi\,,0\le \chi,\psi\le 4\pi$. The Penrose limit is defined by focusing on the geodesic along $\chi$, with $\mu=\pi/4$,
$\alpha=0$, $\rho=0$. To do this we can define the new variables
\begin{equation}
\rho={\tilde \rho\over R},~~~\mu={\pi\over 4} + {u\over 2R},~~~ \alpha = {r\over\sqrt{2} R},~~~
dx^+={dt+d\chi/2\over 2},~~~~
dx^-=R^2{dt-d\chi/2\over 2},
\end{equation}
and scale $R\to \infty$. The metric then reduces to
\begin{equation}
\begin{aligned}
ds^2 &= -4dx^+dx^- + du^2 + d\tilde\rho^2 + \tilde\rho^2 d\Omega_2^2 + dr^2 + {r^2\over 4}\sum_{i=1}^3\sigma_i^2
-(u^2+\tilde\rho^2)(dx^+)^2 +{1\over 2}r^2\sigma_3 dx^+ \\
&= -4dx^+dx^- + du^2 + \sum_{i=1}^3 dy_i^2  + \sum_{j=1}^2 dz_j d\bar z_j
-(u^2+\sum_{i=1}^3 y_i^2)(dx^+)^2 -  {i\over 2}\sum_{j=1}^2 (\bar z_j dz_j - z_j d\bar z_j) dx^+
\end{aligned}
\end{equation}
where $z_1$, $z_2$ are standard complex coordintes on the ${\mathbb C}^2$ with radial coordinates
$(r,\theta,\phi,\psi)$.
To put the metric in standard pp-wave form, we make a further coordinate change
\begin{equation}
z_j = e^{-{i}x^+/2} w_j,~~~~\bar z_j = e^{{i}x^+/2} \bar w_j,
\end{equation}
and the metric becomes
\begin{equation}\label{ppmetric}
\begin{aligned}
ds^2 &= -4dx^+dx^- + du^2 + \sum_{i=1}^3 dy_i^2  + \sum_{j=1}^2 dw_j d\bar w_j
-(u^2+\sum_{i=1}^3 y_i^2 + {1\over 4}\sum_{j=1}^2 |w_j|^2)(dx^+)^2
\end{aligned}
\end{equation}
There are also fluxes in the $AdS_4\times {\mathbb CP}^3$ background, reducing to
\begin{equation}
\begin{aligned}
& F_2 = - dx^+\wedge du, \\
& F_4 = -3 dx^+ \wedge dy_1 \wedge dy_2 \wedge dy_3,
\end{aligned}
\end{equation}
in the Penrose limit.
This pp-wave solution was found in \cite{Bena:2002kq} (see also \cite{Lin:2005nh}),
and preserves 24 supersymmetries
as the $AdS_4 \times {\mathbb CP}^3$ background does.
We shall organize the coordinates $(u,y_i)$ as $(X^1,X^2,X^3,X^4)$,
and $w_i, \bar w_i$ as $(X^5,X^6,X^7,X^8)$.
In the light cone gauge $X^+=\tau$, $\Gamma^+\Theta=0$, the Green-Schwarz action
for the type IIA string is (we follow the conventions of \cite{Michelson:2002ps}; for earlier studies of the GS string in this pp-wave background see \cite{Sugiyama:2002tf},\cite{Hyun:2002wu})
\begin{equation}
\begin{aligned}
S &= {1\over 2\pi\alpha'}\int dt\int_0^{2\pi\alpha' p^+} d\sigma \left\{ {1\over 2}\sum_{i=1}^8 \left[ (\dot X^i)^2 -({X^i}')^2 \right]
-{1\over 2}\sum_{i=1}^4 (X^i)^2 - {1\over 8}\sum_{j=5}^8 (X_j)^2\right. \\
&\left.~~~~~~
-i \bar \Theta\Gamma^-\left[\partial_\tau+\Gamma^{11}\partial_\sigma
-{1\over 4} \Gamma^1\Gamma^{11} -{3\over 4}\Gamma^{234} \right]\Theta
\right\}
\end{aligned}
\end{equation}
The bosonic excitations of the type IIA string in this pp-wave background have light cone
spectrum
\begin{equation}\label{bosexc}
H = \sum_{i=1}^4\sum_{n=-\infty}^\infty N_n^{(i)}\sqrt{1+{n^2\over(\alpha'p^+)^2}}+\sum_{j=5}^8 
\sum_{n=-\infty}^\infty N_n^{(j)}\sqrt{{1\over 4}+{n^2\over (\alpha'p^+)^2}}
\end{equation}
In terms of the gauge theory spin chain variables,
$p^+=J/R^2$,\footnote{To see this, note that
$p^+=-{1\over 2}p_-={i\over 2 R^2}{\partial\over\partial x^-} = {i\over R^2}({1\over 2}\partial_t - \partial_\chi)$.
Since $\chi\sim \chi+4\pi$, the charge quantization is such that $-i\partial_\chi=J/2$, and
$i\partial_t = \Delta$. For the chiral primary with $\Delta=J$ ($J$ is the length of the alternating $A_1B_1$ chain
divided by 2), we have $p^+=J/R^2$.
} $p=n/J$, $R^2/\alpha' = \pi \sqrt{2\lambda}$ \cite{Aharony:2008ug}, we find the dispersion relations
\begin{equation}\label{disp}
\begin{aligned}
&E^{(i)} = \sqrt{1+2\lambda(\pi p)^2},~~~~~i=1,\cdots,4,\\
& E^{(j)}=\sqrt{{1\over 4}+
2\lambda(\pi p)^2},~~~~~j=5,\cdots,8.
\end{aligned}
\end{equation}
It follows from the fermion equation of motion that
\begin{equation}
\begin{aligned}
(\partial_\tau^2-\partial_\sigma^2)\Theta&=\left(\partial_\tau-\Gamma^{11}\partial_\sigma\right)
\left(\partial_\tau+\Gamma^{11}\partial_\sigma\right) \Theta
\\
&=
-\left({1\over 4} \Gamma^1\Gamma^{11} +{3\over 4}\Gamma^{234} \right)^2\Theta\\
&=\left( {5\over 8}+{3\over 8}\Gamma^{1234}\Gamma^{11}\right)\Theta
\end{aligned}
\end{equation}
Hence there are four fermions of mass $1$, satisfying $\Gamma^{1234}\Gamma^{11}\Theta=\Gamma^{5678}\Theta=\Theta$,
and four fermions of mass $1/2$, satisfying $\Gamma^{1234}\Gamma^{11}\Theta=\Gamma^{5678}\Theta=-\Theta$. Consequently the fermion
spectrum takes the same form as the bosonic one (\ref{bosexc}). Note that the Green-Schwarz action has symmetry group
$SU(2)'\times SU(2|2)$, which contains bosonic subgroup $SU(2)'\times SU(2)_G\times SU(2)_r\times U(1)_D$. Here
$SU(2)'\times SU(2)_G\simeq SO(4)$ is the rotation group on $(X^5, X^6, X^7, X^8)$, whereas
$SU(2)_r$ rotates $(X^2, X^3, X^4)$. The supersymmetry transformations of the $X^i$'s take the form
\begin{equation}
\begin{aligned}
& \delta_{A\alpha} u \sim \Theta_{A\alpha},\\
& \delta_{A\alpha} y^i \sim {(\sigma^i)_\alpha}^\beta\Theta_{A\beta},\\
& \delta_{A\alpha} X^{B\dot C} \sim \delta_A^B \Theta^{\dot C}_{\alpha},
\end{aligned}
\end{equation}
where $X^{A\dot B}$ stand for $(X^5, X^6, X^7, X^8)$ in $SU(2)_G\times SU(2)'$ bispinor notation.
This is consistent with the fact that $\Theta_{A\alpha}$ (satisfying $\Gamma^{5678}\Theta=\Theta$)
have the same mass as $(u,y^i)$, and $\Theta_{\dot A\alpha}$ (satisfying $\Gamma^{5678}\Theta=-\Theta$) have the same mass as $X^{A\dot B}$. The
$SU(2)'$ symmetry appears to be an accidental symmetry in the pp-wave limit, and 
reduces to a $U(1)$ away from the Penrose limit.

We shall note an important difference
of this pp-wave limit from say the BMN scaling of ${\cal N}=4$ SYM \cite{Berenstein:2002jq}: the limit here is defined by
taking $\lambda, J\to \infty$, while keeping $\lambda/J^2$ fixed.
This may appear surprising from perturbative gauge theory, as we might have expected from the two-loop dispersion relation
(\ref{disptwo}) that the BMN scaling might be defined by $\lambda/J$ kept fixed.
On other hand, in general the $\ell$-loop corrections may contribute to the
dispersion relation in the form
\begin{equation}
E^{(\ell)} = \lambda^{\ell} \sum_{n=1}^{\lfloor\ell/2\rfloor} c_{\ell,n}\sin^{2n}(\pi p)
\end{equation}
where $c_{\ell, n}$ are generically nonzero (say for $n=1$ and $\ell>2$), and hence alters the form of the BMN scaling
at strong coupling. This indeed seems to happen in ${\cal N}=6$ CSM theory.

At classical dimension $1/2$, there are 4 bosonic and 4 fermionic excitations. They are the
modes of $(X^{A\dot B},\Theta^{\dot B}_\alpha)$,
where $\dot B$ is an $SU(2)'$ spinor index, and lie
in the two short multiplets $(2|2)_A$ and $(2|2)_B$ with respect to $SU(2|2)$.
Their exact dispersion relation at general 't Hooft coupling is given by
the BPS bound (\ref{bound}), where the function $f(\lambda)$ scales differently with $\lambda$
in the weak and strong coupling limits, see (\ref{disptwo}) and (\ref{disp})
\begin{equation}
\begin{aligned}
&f(\lambda) \sim \lambda,~~~~\lambda\ll 1,\\
&f(\lambda) \sim \sqrt{\lambda/2},~~~~\lambda\gg 1.
\end{aligned}
\end{equation}
A similar phenomenon was observed in \cite{Lin:2005nh}. This is in contrast with ${\cal N}=4$ SYM, where the central charge of the extended superconformal algebra
of the infinite spin chain is proportional to $\sqrt{\lambda}$ in both the weak and strong coupling limits
(although there is no reason why this should be true at general finite 't Hooft coupling, as pointed out in \cite{Hofman:2006xt}).

At classical dimension $1$, we have pairs of free excitations in
$(2|2)_A\otimes (2|2)_A$, $(2|2)_B\otimes (2|2)_B$,
$(2|2)_A\otimes (2|2)_B$, as well as an additional $(4|4)$ multiplet of spin content
$\{[0,2],[0,0]|[1,1]\}$ under $SU(2)_G\times SU(2)_r$, which are the modes of $(y^i,u,\Theta_{A\alpha})$.
Note that the dispersion relation of the $(4|4)$ multiplet is consistent with the form of the BPS bound
for $(4|4)$ short multiplets at generic coupling,
\begin{equation}
D = \sqrt{1+4f(\lambda)^2 \sin^2(\pi p)}
\end{equation}
It is plausible that this $(4|4)$ multiplet survives as a short multiplet away from the pp-wave limit.
Naively, we may expect
this multiplet to include the $D_\mu$ impurities. But as we have seen
at two-loop, the $D_\mu$ impurities mix with the $(2|2)_A\otimes (2|2)_B$
sector to form threshold scattering states, and there are no bound states
(at least not at two loop). It is a puzzle to us how to describe this
$(4|4)$ multiplet perturbatively in the gauge theory, if it exists.

In the $(2|2)_A\otimes (2|2)_A$ (or
$(2|2)_B\otimes (2|2)_B$) sector, at two-loop we have found bound states
that saturate the BPS bound; they may survive as short multiplets at finite coupling,
and may become free pairs of $(2|2)_A$ (or $(2|2)_B$) excitations in the pp-wave limit.
In $(2|2)_A\otimes (2|2)_B$ sector, we have found an $8+8$ long multiplet of
threshold (non-)scattering states at two-loop. It is unclear whether these survive at finite coupling,
and match onto the free pairs of $(2|2)_A$ and $(2|2)_B$ excitations in the pp-wave limit.

\section{Giant magnons}

It is easy to find giant magnon
solutions to the Nambu action in $AdS_4\times {\mathbb{CP}}^3$, following \cite{Hofman:2006xt}.
Corresponding to our vacuum spin chain is a string moving along a geodesic in the ${\mathbb{CP}}^3$,
with $\mu=\pi/4$ and $\alpha=0$, parameterized by $\chi$, in the coordinate
system of (\ref{cpmetric}). Alternatively, we can work with projective coordinates
$[z_1,z_2,z_3,z_4]$, and consider the geodesic given by $|z_1|=|z_3|$, $z_2=z_4=0$. The first type of giant
magnons move on the $S^2$ parameterized by $\chi$ and $\mu$, at $\alpha=0$. In projective coordinates,
this is the ${\mathbb{CP}}^1$ defined by $z_2=z_4=0$. Note that this sphere preserves the $SU(2)_G$ which rotates
$z_2$ and $z_4$. In particular it is consistent to restrict the giant magnon solution
to this $S^2$. The $S^2$, or $\mathbb{CP}^1$, has its radius equal to the $AdS_4$ radius $R=\sqrt{\pi\alpha'}(2\lambda)^{1\over 4}$.
The giant magnon solution takes the identical form as the one in \cite{Hofman:2006xt}, with dispersion relation
\begin{equation}
E-J = \sqrt{2\lambda}|\sin({\Delta\chi\over 4})|=\sqrt{2\lambda}|\sin(\pi p)|
\end{equation}
where the angular difference between two ends of the magnon, $\Delta\chi/2$, is identified with $2\pi p$
($\chi$ has periodicity $4\pi$). This is consistent with (and saturates) the large $\lambda$ limit of the BPS bound due to the centrally
extended $SU(2|2)$ algebra of the infinite chain.

Interestingly, there is a second class of giant magnons, which lie in an $\mathbb{RP}^2\subset \mathbb{CP}^3$,
rather than the ${\mathbb{CP}}^1$. The $\mathbb{RP}^2$'s that contain the geodesic $|z_1|=|z_3|$, $z_2=z_4=0$
are defined by
\begin{equation}
|z_1|=|z_3|,~~~~{z_2\over z_1-z_3}=\alpha x,~~~~{z_4\over z_1-z_3}=\beta x,~~~~x\in\mathbb{R}.
\end{equation}
This is seen more explicitly in rotated coordinates $(\tilde z_1,\tilde z_2,\tilde z_3,\tilde z_4) =
({z_1-z_3\over\sqrt{2}},z_2, i{z_1+z_3\over \sqrt{2}},z_4)$, where
the original geodesic is $\tilde z_3/\tilde z_1\in \mathbb{R}$, $\tilde z_2=\tilde z_4=0$,
and the defining equation of the ${\mathbb{RP}}^2$ becomes
\begin{equation}
{\tilde z_3\over \tilde z_1}\in {\mathbb R},~~~~{\tilde z_2\over \tilde z_1}=\alpha x,~~~~{\tilde z_4\over \tilde z_1}=\beta x,~~~~x\in\mathbb{R}.
\end{equation}
where $\alpha$ and $\beta$ are complex constants, and we have a family of $\mathbb{RP}^2$'s parameterized by $(\alpha,\beta)$, which
transform as a doublet under $SU(2)_G$. In particular, a given $\mathbb{RP}^2$ lies in a $\mathbb{CP}^2\subset \mathbb{CP}^3$
which is fixed by a $U(1)$ symmetry, and it is the fixed locus of an involution of the $\mathbb{CP}^2$. It is therefore consistent
to restrict the giant magnon solution to this $\mathbb{RP}^2$.
We can describe the $\mathbb{RP}^2$ as the quotient of an auxiliary sphere $\tilde S^2$ by
the antipodal map. Note however that this $\tilde S^2$ has radius $2R$. Hence the giant magnon solutions
that move along the geodesic, which is half the equator of $\tilde S^2$ (with ends identified by the antipodal map),
has dispersion relation
\begin{equation}
E-J = 2\sqrt{2\lambda}|\sin({\Delta\varphi\over 2})|
\end{equation}
where $\varphi$ is the angular variable on the equator of $\tilde S^2$, ranging from 0 to $2\pi$, and
$\Delta\varphi$ is the difference between the two ends of the giant magnon. On the $\mathbb{RP}^2$, however, $\varphi$
is identified with periodicity $\pi$, and it is natural to propose the identification with spin chain momentum
$\Delta\varphi = \pi p$. So we obtain the dispersion relation
\begin{equation}
E-J = 2\sqrt{2\lambda}|\sin(\pi p/2)|
\end{equation}
Note that with given $0<p<1$, there is another giant magnon with $\Delta\varphi = \pi(1-p)$ with the same ends
as the one with $\Delta\varphi = \pi p$, and has dispersion relation $E-J=2\sqrt{2\lambda}|\cos(\pi p/2)|$. The minimal energy
configuration carrying momentum $p$ should then be
\begin{equation}
E-J = 2\sqrt{2\lambda} \min\{|\sin(\pi p/2)|,|\cos(\pi p/2)|\}
\end{equation}
Note that this obeys the large $\lambda$ limit of the BPS bound (\ref{bound}), but does not saturate it.

Naively, based on the transformation under $SU(2)_G$, one may want to identify the first type of giant magnons
with the $(4|4)$ multiplet in the pp-wave limit, since it involves excitations in the $u$-direction (see
(\ref{ppmetric})), and to identify
the second type of giant magnons with the $(2|2)$ multiplets. However, the second type of giant magnons does not saturate
the BPS bound of the $SU(2|2)$ algebra, and should correspond to long multiplets. A potential resolution to this puzzle
is that there are fermion zero modes of the giant magnons, which carry additional representations of the $SU(2)_G$. It is therefore
not clear to us how to identify these giant magnons with the excitations in the pp-wave limit or in perturbative gauge theory.

\subsection*{Acknowledgments}

We are grateful to H. Lin, J. Maldacena,
A. Tomasiello and E. Witten for useful discussions, and especially to D. Jafferis for sharing with us an early draft.
The work of D.G. is supported in part by DOE grant DE-FG02-90ER40542.
The work of S.G. is supported in part by the Center for the
Fundamental Laws of Nature at Harvard University and by NSF grants PHY-024482 and DMS-0244464.
The work of X.Y. is supported by a Junior Fellowship from the Harvard Society of Fellows. S.G. thanks the 6th Simons Workshop in Mathematics and Physics at Stony Brook for hospitality during completion of this work.

\end{document}